\documentclass[lettersize,journal]{IEEEtran}
\usepackage{amsmath,amsfonts}
\usepackage{algorithmic}
\usepackage{algorithm}
\usepackage{array}
\usepackage[caption=false,font=normalsize,labelfont=sf,textfont=sf]{subfig}
\usepackage{textcomp}
\usepackage{stfloats}
\usepackage{url}
\usepackage{verbatim}
\usepackage{graphicx}
\usepackage{cite}
\usepackage{booktabs}
\usepackage{multirow}
\usepackage{tabularx}
\usepackage{bbding}

\usepackage{titlesec}
\titlespacing{\section}{0pt}{0.6\baselineskip}{0.2\baselineskip}
\titlespacing{\subsection}{0pt}{0.3\baselineskip}{0.2\baselineskip}
\usepackage{color}

\usepackage{nccmath}
\begin{document}

\title{MSV-Mamba: A Multiscale Vision Mamba Network for Echocardiography Segmentation}

\author{Xiaoxian Yang,
        Qi Wang,
        Kaiqi Zhang,
        Ke Wei,
        Jun Lyu*,
        Lingchao Chen*
        
\thanks{This work was supported by the National Natural Science Foundation of China under Grant 61902236. (Corresponding authors: Jun Lyu; Lingchao Chen.)}

\thanks{Xiaoxian Yang, Qi Wang, Kaiqi Zhang, and Ke Wei are with the School of Computer and Information Engineering, Shanghai Polytechnic University, Shanghai 200120, China (e-mail: xxyang@sspu.edu.cn; 20221513011@stu.sspu.edu.cn; 20231513051@sspu.edu.cn; 20231513045@stu.sspu.edu.cn).}


\thanks{Jun Lyu is with the Department of Clinical Research, The First Affiliated Hospital of Jinan University, Guangzhou, China (e-mail: lyujun2020@jnu.edu.cn).}

\thanks{Lingchao Chen is with the Department of Neurosurgery, Fudan University Huashan Hospital, China (e-mail: chenlingchao12@163.com).}

}



\maketitle

\begin{abstract}
Echocardiographic image segmentation plays a crucial role in analyzing cardiac function and diagnosing cardiovascular diseases. Ultrasound imaging frequently encounters challenges, such as those related to elevated noise levels, diminished spatiotemporal resolution, and the complexity of anatomical structures. These factors significantly hinder the model's ability to accurately capture and analyze structural relationships and dynamic patterns across various regions of the heart. Mamba, an emerging model, is one of the most cutting-edge approaches that is widely applied to diverse vision and language tasks. It efficiently captures global information with linear complexity and compensates for the shortcomings of convolutional neural networks (CNNs) and conventional transformers. To this end, this paper introduces a U-shaped deep learning model incorporating a large-window Mamba scale (LMS) module and a hierarchical feature fusion approach for echocardiographic segmentation. First, a cascaded residual block serves as an encoder and is employed to incrementally extract multiscale detailed features. It addresses the vanishing gradient issue by leveraging a residual structure that ensures stable and rapid convergence throughout the training process. Second, a large-window multiscale mamba module is integrated into the decoder to capture global dependencies across regions and enhance the segmentation capability for complex anatomical structures. Furthermore, our model introduces auxiliary losses at each decoder layer and employs a dual attention mechanism to fuse multilayer features both spatially and across channels. This approach enhances segmentation performance and accuracy in delineating complex anatomical structures. Finally, the experimental results using the EchoNet-Dynamic and CAMUS datasets demonstrate that the model outperforms other methods in terms of both accuracy and robustness. For the segmentation of the left ventricular endocardium (${LV}_{endo}$), the model achieved optimal values of 95.01 and 93.36, respectively, while for the left ventricular epicardium (${LV}_{epi}$), values of 87.35 and 87.80, respectively, were achieved. This represents an improvement ranging between 0.54 and 1.11 compared with the 
best-performing model.

\end{abstract}

\begin{IEEEkeywords}
echocardiography segmentation, Mamba, feature fusion, residual block, cardiovascular disease diagnosis.
\end{IEEEkeywords}

\vspace{20 mm}
\section{Introduction}
\IEEEPARstart
Heart disease remains a leading cause of mortality worldwide. According to the World Health Organization (WHO), cardiovascular diseases account for more than 17 million deaths annually, representing more than 30$\%$ of the global death toll \cite{1-townsend2022epidemiology, 2-zhao2021epidemiological}. The functional analysis of the left ventricle (LV) is particularly important for cardiac assessment. As the central component of the heart's pumping mechanism, the left ventricle is essential for maintaining circulatory function, and its structural or functional abnormalities are closely linked to numerous severe cardiac conditions \cite{3-halliday2021assessing, 37-yildiz2020left, 40-nagueh2020left}. Consequently, precise segmentation of the left ventricle is of great clinical importance. It helps clinicians assess cardiac systolic and diastolic function more accurately, which facilitates rapid and reliable diagnoses of conditions such as myocardial infarction and heart failure.

Echocardiography is a frequently employed imaging modality for assessing heart function \cite{38-pellikka2020guidelines, 39-pandian2023recommendations}. Its widespread application in clinical cardiology is attributed to its key advantages, including its noninvasiveness, cost-effectiveness, and real-time imaging capabilities. In addition to offering two-dimensional visualizations of both the two-chamber and four-chamber views, echocardiography captures dynamic changes over time through time series data \cite{4-zamzmi2022real}. However, ultrasound imaging is inherently limited by high noise levels, low spatial resolution, and blurred regional boundaries. These issues are particularly pronounced in patients with heart disease, where the cardiac anatomy frequently exhibits greater complexity and variability. The manual segmentation of critical regions is a labor-intensive and time-consuming process that requires specialized expertise. It is also prone to operator-dependent variability, which ultimately compromises the accuracy and consistency of the results. Therefore, the development of automated methods for echocardiographic segmentation is needed.

In recent years, deep learning methods have achieved remarkable progress in medical image segmentation, particularly with the emergence of U-Net. U-Net \cite{5-weng2021inet} employs an encoder--decoder architecture to efficiently integrate local and global information by progressively extracting high-level features with the encoder and restoring spatial resolution with the decoder. The network's skip connection mechanism enables the fusion of features at different levels, leading to robust performance across diverse anatomical structures, such as the liver, kidneys, and brain. With these advancements, numerous model variants based on convolutional neural networks (CNNs) and transformers have been developed. CNNs \cite{6-li2021survey} are effective at capturing intricate image details and segmenting organs such as the heart, lungs, and stomach. Nevertheless, their limited receptive field poses challenges in modeling long-range dependencies, which are critical in scenarios with complex anatomical structures or interorgan relationships. For example, while CNNs precisely delineate tumor boundaries in liver images, they struggle to capture spatial relationships between organs in global structural analysis. In contrast, transformers \cite{7-dosovitskiy2020vit}, with their self-attention mechanism, excel in global feature extraction and interorgan relationship identification of multimodal image analysis. However, their ability to capture fine-grained details is relatively weak, and their computational complexity, scaling quadratically with input size (O(n²)), affects their real-time performance. For example, in brain image segmentation, transformers model intricate relationships between brain regions but may have limitations in detecting subtle anatomical variations. In complex procedures such as echocardiography, these models encounter intrinsic challenges, such as significant computational load and inadequate integration of the global insights and deep-level features. These challenges are obvious in left ventricular segmentation, where achieving high accuracy remains an influential bottleneck. Recently, a visual model incorporating the Mamba module \cite{8-zhu2024vision} that yields excellent results in many tasks was proposed. This model offers a more efficient solution by capturing global features while reducing the computational complexity to a linear scale. This advancement addresses key challenges for improving segmentation performance in echocardiographic imaging.

To address these challenges, this paper presents a U-shaped echocardiogram segmentation model based on multiscale large-window Mamba feature fusion to increase the accuracy and efficiency of left ventricular segmentation. First, cascaded residual blocks are utilized as encoders that a deep network structure leverages to extract high-level image features. Incorporating residual connections mitigates the vanishing gradient problem, ensuring efficient feature learning even at deeper layers. Second, a large-window Mamba scale (LMS) block is introduced as a decoder module to patch and perform pixel-based feature extraction, increasing the model's ability to capture global information. The computational complexity is reduced by the use of a bidirectional state space model. Moreover, an auxiliary loss function is designed for each decoder output layer to ensure that the features learned at each layer contribute more significantly to the final segmentation result. Finally, in the lower three layers of the decoder, multilayer fusion of the dual attention mechanism is incorporated to enhance the model's ability to capture critical features through feature fusion, leading to improved segmentation accuracy. The main contributions of this paper are summarized as follows:

1. The LMS block is introduced as the central module of the decoder. It ensures linear computational complexity while enabling the model to capture global features. The information flow is optimized, and the segmentation performance is improved, particularly in complex backgrounds.

2. Feature fusion is implemented across decoders at different layers using a dual attention mechanism. Spatial and channel attention is applied to multilayer features to prioritize key features and ensure the effective capture of critical information.

3. A hierarchical auxiliary loss function as a learnable strategy is proposed to facilitate collaborative learning across decoder layers. This approach ensures that the output returned from each layer positively influences the final segmentation result; therefore, the overall segmentation accuracy of the model is superior.

The remainder of this paper is structured as follows: Section \uppercase\expandafter{\romannumeral2} reviews related work; Section \uppercase\expandafter{\romannumeral3} details the proposed method; Section \uppercase\expandafter{\romannumeral4} presents experimental results and analysis; and Section \uppercase\expandafter{\romannumeral5} concludes the paper with a summary and future directions.

\section{Related work}
Medical image segmentation is a fundamental task within medical image analysis and holds considerable importance for early disease diagnosis, treatment planning, and clinical research \cite{9-liu2021review}. Owing to the rapid progress in medical imaging technologies such as computed tomography (CT), magnetic resonance imaging (MRI), and ultrasound, the volume of image data accessible for clinical applications has notably increased. However, accurately extracting the contours of target regions, such as tumors or organ boundaries, from these complex images remains a significant challenge in medical image processing. Therefore, medical image segmentation serves not only as a basic component of image analysis but also as a critical aspect of automated medical diagnosis.   

In recent years, deep learning-based methods for medical image segmentation have made notable advancements, with CNNs and their variants emerging as mainstream tools. Since its introduction in 2015, the U-Net model has achieved impressive performance in segmenting a variety of medical images, including CT, MR, and ultrasound images. U-Net effectively captures multiscale features with its symmetrical encoder--decoder structure and combines high- and low-level information through skip connections \cite{5-weng2021inet,6-li2021survey}. Several studies have proposed enhancements to U-Net. For example, the attention U-Net, introduced by Zhu et al. \cite{10-zhu2022attention}, incorporates an attention mechanism to improve segmentation accuracy in complex regions, and it is well suited for complex medical images. Kushnure et al. \cite{11-kushnure2021ms} focused on multiscale feature extraction, and the representation of global and local information was refined, improving segmentation performance while reducing computational complexity and the number of model parameters. Furthermore, Zhang et al. \cite{12-zhang2020attention} introduced AGResU-Net, which integrates residual modules and attention-gated mechanisms into the U-Net structure. 
The attention gates in the skip connections highlight important features such as semantic extraction features, and the model's ability to detect brain tumors is enhanced. 
Feng et al. \cite{13-feng2020ssn} designed a ladder network (SSN) for real-time polyp segmentation in colonoscopy images. In this model, spatial features are extracted during the encoder stage, whereas dual attention and multiscale fusion modules are integrated during the decoder stage. This modification improves the segmentation accuracy, accelerates the inference speed, and outperforms U-Net in real-time applications. These studies demonstrate that architectural innovations and feature optimization in deep learning methods continually refine medical image segmentation performance, addressing diverse clinical needs more effectively. 

Although CNNs are excellent at capturing local features and image details, their limited receptive field restricts their ability to model long-range dependencies. Moreover, the transformer model has attracted attention in medical image segmentation because of its superior capabilities in global feature modeling \cite{7-dosovitskiy2020vit}. Unlike traditional convolutional networks, transformers utilize a self-attention mechanism to capture long-range dependency information, and their ability to represent complex structures is advanced. The TransUNet model proposed by Chen et al. \cite{14-chen2024transunet} integrates transformers with CNN feature maps. This model segments images into patches for label encoding, extracting global context and fusing it with high-resolution CNN features through a decoder to achieve precise positioning. Cao et al. \cite{15-cao2022swin} developed a model based on the Swin transformer, where a hierarchical structure of shifted windows is employed in the encoder to capture contextual features. This model restores spatial resolution through a symmetric decoder with patch expansion layers, enabling enhanced multiorgan and heart segmentation. Pham et al. \cite{16-pham2024seunet} introduced the seUNet-Trans model, which integrates U-Net features with the transformer module through a bridge layer. This model integrates U-Net's spatial positioning capabilities with the transformer's long-range dependency modeling, eliminating the requirement of traditional position embedding. Song et al. \cite{17-song2023tgdaunet} designed the TGDAUNet model, which uses a dual-branch backbone network of a CNN and a transformer. This approach reduces multiscale redundancy through a polarized self-attention module and refines boundaries with a reverse graph reasoning module. Furthermore, Bi et al. \cite{18-bi2023bpat} devised the BPAT-UNet model for segmenting ultrasound thyroid nodules. This model integrates a boundary point supervision module (BPSM) and an adaptive multiscale feature fusion module (AMFFM) to improve boundary features and seamlessly combines features across various scales and channels. These studies highlight the potential of combining a transformer with a CNN, which increases the segmentation performance for complex medical images and contributes to the advancement of real-time medical applications.

However, the computational complexity of the transformer model is high, which may reduce real-time performance when processing high-resolution medical images. To address this issue, researchers have proposed various lightweight modifications to reduce computational overhead while preserving global feature modeling capabilities and augmenting the model's practical applicability in medical image analysis. The visual Mamba model \cite{8-zhu2024vision} represents one such innovation. This model offers new possibilities for real-time analysis by simplifying the computational structure. Ruan et al. \cite{19-ruan2024vm} introduced the VM-UNet model to operate the visual state space (VSS) block as a core component for capturing an extensive range of contextual information. It further establishes an asymmetric encoder--decoder architecture with exceptional performance. Liao et al. \cite{20-liao2024lightm} integrated the Mamba module into the U-Net framework to propose the LightM-UNet model. This model leverages a residual visual Mamba layer to perform deep semantic feature extraction and model long-range dependencies, all while maintaining linear computational complexity. Wang et al. \cite{21-wang2024mamba} designed the Mamba-UNet model, which integrates jump connections into the VMamba-based encoder--decoder structure to preserve spatial information across different scales. This approach enables comprehensive feature learning and effective capture of complex details. Moreover, the large kernel visual mamba network (LKM-UNet) proposed by Wang et al. \cite{22-wang2024lkm} outperforms traditional CNN and transformer architectures in local spatial modeling. This model maintains favorable global feature modeling capabilities through the optimization of the self-attention mechanism. Further studies by Wang et al. \cite{23-wang2024weak} led to the development of the weak-Mamba-UNet model, which incorporates three distinct encoder--decoder networks: a CNN-based U-Net for local feature extraction, a SwinUNet based on the Swin transformer for global context modeling, and a Mamba-UNet for efficient long-range dependency modeling. This model adopts a collaborative and cross-supervision mechanism to achieve iterative learning and segmentation refinement through pseudolabeling. These mamba-based segmentation models present considerable potential in terms of the efficiency and accuracy of complex medical image tasks and offer promising opportunities for real-time clinical applications.

Although deep learning models have made meaningful advances in medical image segmentation, several challenges remain, especially in echocardiography segmentation. Noise and artifacts usually limit precise delineation of boundary regions, and interpatient variability in cardiac anatomical structures and motion patterns limits generalizability. The current segmentation models leveraging Mamba have demonstrated significant potential in image segmentation tasks because they extract global features with linear complexity. This approach compensates for the limitations of the CNN and transformer models. This paper introduces a segmentation model designed for precise segmentation of echocardiograms via vision Mamba to aid in the detection and diagnosis of heart diseases.

\section{Method}

In this section, a multiscale vision mamba network model (MSV-Mamba) for echocardiography segmentation is introduced to improve the performance of ultrasound image segmentation tasks. The model architecture, illustrated in Figure 1, comprises several key modules: an encoder, a decoder, a multilayer feature fusion module, and an auxiliary loss function.

\begin{figure*}[!htb]
\centering
\includegraphics[width=0.96\textwidth]{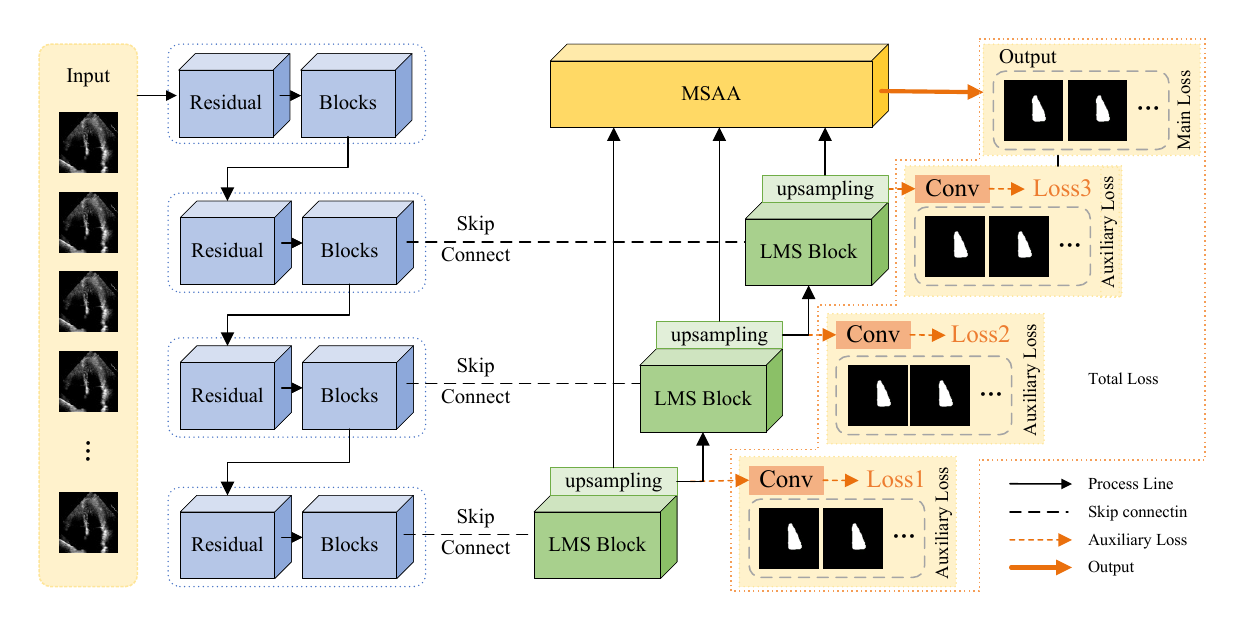}
\caption{Overall structure of the MSV-Mamba model.}
\label{fig_1}
\end{figure*}

The encoder in the proposed model is composed of four cascaded residual blocks, each comprising skip connections, to mitigate the vanishing gradient issue and facilitate efficient feature transfer. As illustrated in Figure 2, each residual block integrates convolutional layers, batch normalization, and rectified linear unit (ReLU) activation functions. This design improves the model's nonlinear expression capability and stability of the training process. As the spatial dimension of the feature map progressively decreases while the number of feature channels increases, the encoder extracts more comprehensive contextual information to establish a robust basis for subsequent decoding. Furthermore, residual connections promote information flow within the network and capture intricate features. This structural design allows the encoder to preserve vital feature information when processing high-dimensional data, enhancing the overall model performance and accuracy.

The decoder leverages the LMS module to strengthen its ability to learn global features. Each LMS block performs feature extraction at the patch and pixel levels, enabling the model capture global context and fine details of ultrasound images more effectively. The decoder progressively restores the spatial resolution of the image through layer-by-layer upsampling. After each upsampling operation, the decoder incorporates skip connections to merge feature maps from the corresponding encoder layers and then starts to address the next layer. Detailed spatial information is preserved by combining both deep and shallow features. This structural design helps the model handle diverse challenges and achieve good overall performance when processing complex ultrasound images.

\begin{figure}[!htb]
\centering
\includegraphics[width=0.45\textwidth]{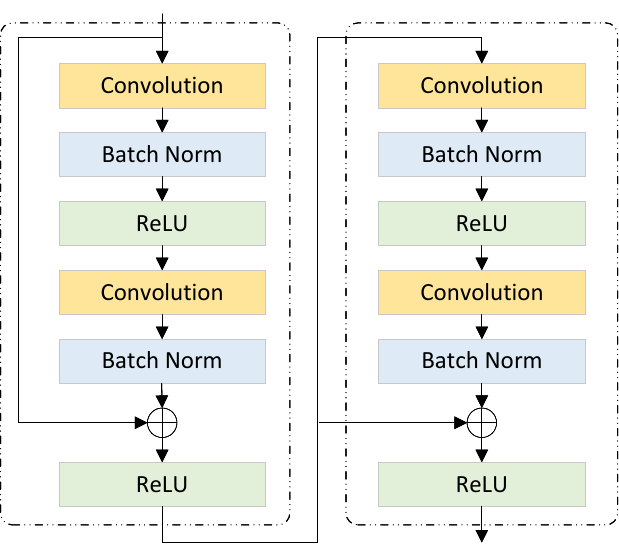}
\caption{Residual block structure diagram.}
\label{fig_2}
\end{figure}

At the highest level of the decoder, the multiscale attention aggregation (MSAA) module is employed to achieve spatial and channel-level fusion of multilayer features. The global context and local details are incorporated through weighted feature fusion across various layers, enhancing the model's ability to learn complex structures. The MSAA module not only captures feature details across multiple scales but also improves the model's segmentation precision. This fusion approach provides the model with a holistic perspective, ensuring that subtle anatomical structures and vital functional attributes are accurately represented during cardiac image processing.

To improve the contribution of each decoder layer to the final mask prediction, we incorporate an auxiliary loss function after the upsampling operation of each LMS block. This design enables the influential features of each layer during training to be captured and utilized while promoting synergy among features at various levels. Supplementary supervision is applied to each decoder layer during training, enabling the model to learn the nuanced distinctions and relevance of features at every layer and thus optimizing the final segmentation outcome.

\subsection{Large-window Mamba scale module}
LMS blocks serve as the core components of the decoder and are designed to spatially model multiscale feature maps at each stage, allowing more precise cardiac image segmentation to be achieved. As shown in Figure 3, each LMS block integrates a pixel-level spatial state module (PiM) and a patch-level spatial state module (PaM). The PiM focuses on features extracted from local pixel neighborhoods, whereas the PaM captures global long-range dependencies. This dual-level design is leveraged to detect subtle cardiac changes, such as variations in valve motion and ventricular wall thickness, and identify interrelationships between different anatomical structures of the heart. These findings support the model's ability to achieve more precise segmentation of critical cardiac regions and more credible delineation of regional boundaries.

When feature ${{F}_{l}}$ is input, the feature map is initially partitioned into multiple larger subwindows ${{V}_{l}}$ using a pixel-level spatial similarity module (SSM). For example, given an input feature map with a resolution of $H \times W$, each feature map is divided into subwindows of size $m \times n$. Assuming that $H/m$ and $W/n$ are integers, the feature map is subdivided into $(H\times{W})/(m\times{n})$ subwindows. These subwindows are subsequently processed by the Mamba layer, where local neighboring pixels are iteratively fed into the SSM. This approach models the relationships among local neighborhood pixels that facilitate the identification of ventricular wall continuity and valve motion. Thus, the model preserves critical anatomical details during segmentation. The process is formalized as follows:

\begin{equation}
\label{formula (1)}
{{{F}'}_{l}}=PiM({{F}_{l}})
\end{equation}

\begin{figure}[!htb]
\centering
\includegraphics[width=0.46\textwidth]{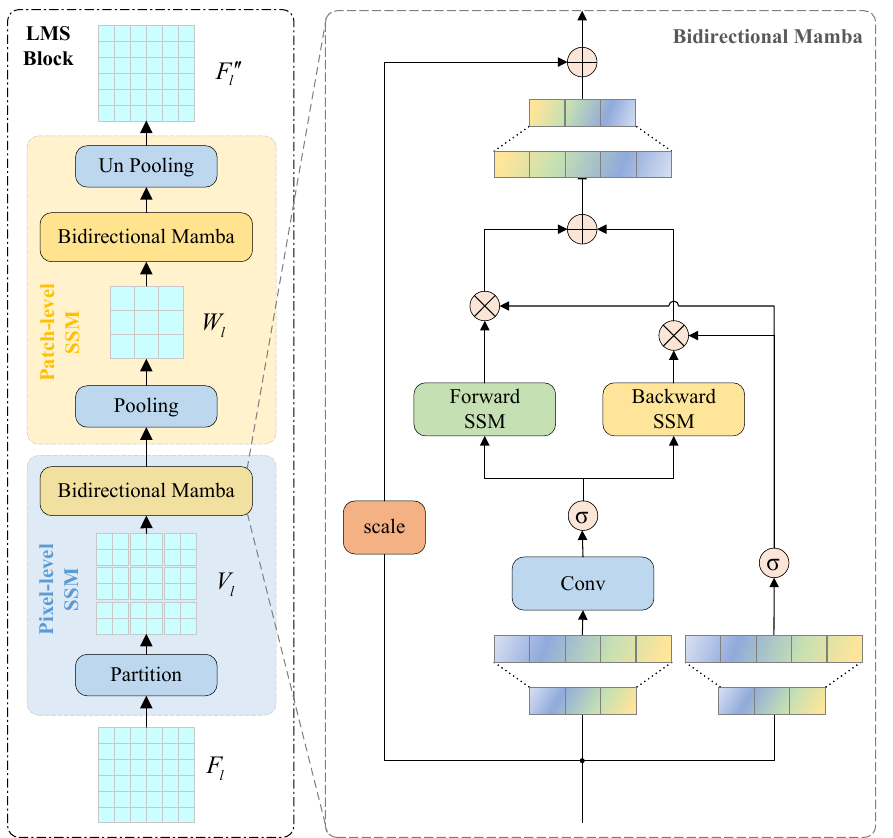}
\caption{LMS block structure diagram.}
\label{fig_3}
\end{figure}

Next, a large-window strategy is implemented to extend the receptive field and capture finer details within ultrasound images. Since these subwindows are nonoverlapping, an effective mechanism is required to facilitate information transfer across windows, enabling the modeling of long-range dependencies between different cardiac regions. Therefore, a patch-level SSM layer is introduced. First, the feature map undergoes processing by a pooling layer of size $m\times{n}$ that summarizes the essential information within each subwindow and generates an aggregate map comprising $(H\times{W})/(m\times{n})$ representative features. The modeling of the overall structure of the heart is improved. Second, these representative features are passed through the Mamba layer, allowing for interactions between subwindows to capture global dependencies. Finally, the postinteraction aggregate map is unpooled back to its original resolution to ensure detailed reconstruction. The process is formalized as follows:


\begin{equation}
\label{formula (2)}
\begin{split}
{{{F}''}_{l}} &= PaM({{{F}'}_{l}}) \\
&= UnPooling(BiMamba(Pooling({{{F}'}_{l}})))
\end{split}
\end{equation}

\begin{equation}
\label{formula (3)}
{{F}_{l+1}}=upSampling({{{F}''}_{l}})
\end{equation}

In this process, $Pooling$ and $Unpooling$ represent the pooling and unspooling layer operations, respectively. $BiMamba$ denotes the bidirectional structure employed in each SSM layer of the LMS block, including the pixel- and patch-level SSMs. This bidirectional design includes simultaneous forward and reverse scanning. Then, the output results are combined to enhance feature modeling capabilities. $upSampling$ refers to progressively upsampling the output results to restore the feature maps to their original spatial resolution. In addition, a scaling factor, $scale$, is introduced in the residual connection to adjust the contribution of features, further optimizing the model performance. The process is formalized as Formula \ref{formula (4)}:

\begin{equation}
\label{formula (4)}
{{F}_{out}}=Bim({{F}_{in}})+scale\cdot {{F}_{in}}
\end{equation}

In the LMS block, the model's ability to focus on the central regions of the image is enhanced, enabling the identification of patches containing more critical information. Moreover, it captures both absolute and relative positional relationships within each patch while establishing robust connections between patches; therefore, the feature representation capabilities of the model are improved. Through precise analysis of complex cardiac images, a reliable foundation for clinical applications is provided that is beneficial for assessing cardiac function and diagnosing disease states.

\subsection{Multiscale attention aggregation module for key regions in echocardiography}

In the MSAA module, a feature fusion approach is designed to integrate the output features from different layers of the decoder, such as ${F}{i}$, ${F}{i-1}$, and ${F}_{i+1}$, to provide the model with richer information. As illustrated by Formula \ref{formula (5)}, a combined feature map is generated through a connection operation and then fed into the MSAA module for image feature fusion. The structure of the MSAA module is shown in Figure 4. This module enhances the model's feature representation capacity by capturing rich anatomical details at multiple levels. This improved feature integration facilitates more accurate identification of the structure and function of the heart.

\begin{equation}
\label{formula (5)}
{{\hat{F}}_{i}}=Concat({{F}_{i}},{{F}_{i-1}},{{F}_{i+1}})
\end{equation}

\begin{figure*}[!htb]
\centering
\includegraphics[width=0.8\textwidth]{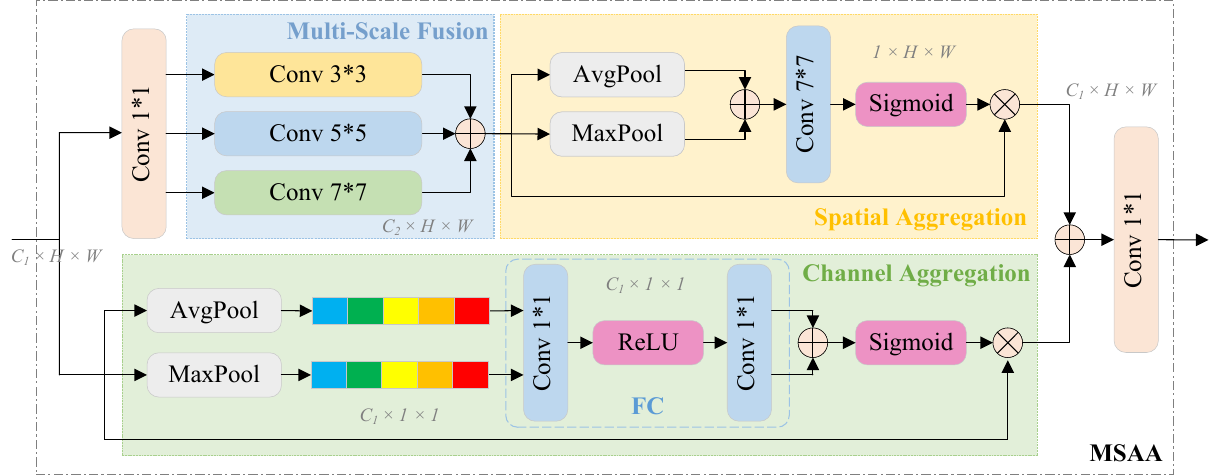}
\caption{MSAA module diagram.}
\label{fig_4}
\end{figure*}

In the feature aggregation process of this module, a dual path is designed to extract spatial and channel features simultaneously. For spatial feature extraction, the number of channels in ${\hat{F}}_{i}$ is initially reduced from $C_{1}$ to $C_{2}$ using a $1\times1$ convolution layer to obtain ${F}_{1}$. The computational complexity is reduced while key cardiac anatomical features are preserved. Feature fusion is then performed through multiscale convolution operations with $3\times3$, $5\times5$, and $7\times7$ kernels to obtain ${F}_{2}$, thus capturing image features at various scales. This process is formalized as follows:

\begin{equation}
\label{formula (6)}
F_{1} = {Conv1}({\hat{F}}_{i})
\end{equation}

\begin{equation}
\label{formula (7)}
F_{2} = {Conv3}(F_{1}) + {Conv5}(F_{1}) + {Conv7}(F_{1})
\end{equation}

After the above operation, the ventricular wall's local texture features and the heart's global morphology are captured and analyzed. A more representative feature map is then generated by merging average pooling and maximum pooling operations to aggregate critical information from different regions. Finally, these feature maps are combined with sigmoid-activated feature maps and processed by $7\times7$ convolutions and elementwise multiplication to enhance nonlinear feature representations. As shown in Formula \ref{formula (8)}, the model better captures subtle anatomical relationships between different heart structures.

\begin{equation}
\label{formula (8)}
\small
F_{spatial} = {Conv7}({AvgPool}(F_{2}) + {MaxPool}(F_{2})) \odot \sigma(F_{2})
\end{equation}

In channel aggregation, the MSAA module first decreases the feature map dimensions to $C1\times1\times1$ by global average pooling and maximum pooling operations and then summarizes the most salient features within the overall context. Afterward, a channel attention map is generated through a $1\times1$ convolution followed by a ReLU activation function to emphasize key features pertinent to specific cardiac regions, such as the left ventricle. This channel attention map is then expanded to match the original input size to fuse with the spatial feature map. The model's sensitivity to critical cardiac regions is amplified, facilitating accurate identification and segmentation of intricate cardiac structures. This process is formalized as follows:

\begin{equation}
\label{formula (9)}
\small
F_{3} = {GlobalAvgPool}({\hat{F}}_{i}) + {GlobalMaxPool}({\hat{F}}_{i})
\end{equation}

\begin{equation}
\label{formula (10)}
\small
F_{channel} = {Sigmoid}({FC}({ReLU}({Conv1}(F_{3}))))
\end{equation}

Finally, the spatial feature $F_{spatial}$ and the channel feature $F_{channel}$ are fused to obtain the feature vector ${\hat{F}}$ that incorporates multiscale spatial and channel attention features. This dual-path feature fusion approach strengthens the model's representation capabilities in the spatial and channel domains, increasing the accuracy of complex cardiac structure recognition.

\subsection{Auxiliary loss}

In this section, an intermediate supervision mechanism is introduced at each LMS block to gradually direct the decoder to produce a semantic segmentation map that contains crucial anatomical information. Through layerwise optimization, this mechanism has a positive effect on the segmentation results at each training stage; therefore, the model's ability to capture complex cardiac structures is reinforced. Specifically, the intermediate output of the LMS block at the i-th layer is formalized as:

\begin{equation}
\label{formula (11)}
{{p}^{i}}=Conv(F_{cs}^{i})
\end{equation}

Let $F_{cs}^{i}$ denote the feature map of the i-th layer LMS block. The feature map is transformed into an output prediction map with $C$ channels by a convolutional module, corresponding to the class probabilities. The overall loss function used for network training is expressed as follows:


\begin{equation}
\label{formula (12)}
\begin{split}
{{L}_{main}} &= {JointLoss(XCELoss(logit}{{{s}}_{{main}}}{,labels),} \\
& \quad {DiceLoss(logit}{{{s}}_{{main}}}{,labels))}
\end{split}
\end{equation}

\begin{equation}
\label{formula (13)}
L_{aux}^{i}={XCELoss(logits}_{{aux}}^{{i}}{, labels)}
\end{equation}

\begin{equation}
\label{formula (14)}
Loss={{L}_{main}}+\varepsilon \cdot \sum\nolimits_{i=1}^{n}{{{\omega }_{i}}*L_{aux}^{i}}
\end{equation}

Among the loss function, $XCELoss$ applies distinct cross-entropy loss functions on the basis of different tasks. For singular segmentation tasks, the binary cross-entropy loss function is chosen, whereas for multisegmentation tasks, the cross-entropy loss function is employed. $DiceLoss$ represents the Dice loss function. In the primary loss function ${L}_{main}$, a combination of $XCELoss$ and $DiceLoss$ is employed to optimize the model's segmentation performance. Specifically, cross-entropy loss measures the model's classification accuracy among diverse pixel categories and discriminates cardiac images concerning various anatomical structures, such as ventricles, atria, and valves. The $Dice$ loss focuses on the overlap between the predicted segmentation and the true region and is particularly crucial for handling imbalanced classes. This approach addresses small or ambiguous target regions, such as left ventricular hypertrophy or endocardial thickening. The main loss function balances classification accuracy and segmentation quality by integrating these two loss functions; therefore, the model's overall performance is improved.

To fully leverage the information from different network layers, a learnable weighted cross-entropy loss function is introduced, enhancing the contribution of each network level to the model's segmentation performance. As shown in Formulas 8 and 9, the auxiliary loss function employs $XCELoss$. The influence of each layer on the final output is dynamically adjusted by learnable weights incorporated into the auxiliary loss. During the training process, a hierarchical supervision method is employed to benefit from the advantages of shallow and deep features. Shallow features mainly capture detailed information such as ventricular wall texture, whereas deep features are more conducive to the analysis and positioning of overall heart morphology. Through this hierarchical supervision approach, the critical information in key regions is continuously optimized and utilized to provide stronger support for echocardiogram segmentation.

\section{Experiments}
\subsection{Dataset}
EchoNet-Dynamic\footnote{https://echonet.github.io/dynamic/index.html} is the primary dataset utilized in this work. This dataset comprises 10,030 two-dimensional echocardiographic recordings of independent individuals collected at Stanford University Hospital between 2016 and 2018 \cite{30-ouyang2019echonet}. As one of the largest publicly available echocardiographic datasets, EchoNet-Dynamic offers comprehensive annotations and represents a diverse patient population. Each video frame has a resolution of $112\times112$ pixels, and detailed information is provided for the end-diastolic volume, end-systolic volume, left ventricular contour, and ejection fraction associated with each video. The dataset is divided into 7,465 training samples, 1,288 validation samples, and 1,277 test samples. The training set exhibits a data imbalance, with only 12.7$\%$ of the samples having an EF ratio below 40$\%$.

Experiments were also conducted using the CAMUS\footnote{https://www.creatis.insa-lyon.fr/Challenge/camus/} dataset, which comprises cardiac ultrasound data from 500 patients at the Saint-Etienne University Hospital in France \cite{32-leclerc2019deep}. Among them, 450 patients were designated for the training set, and 50 new patients were allocated to the test set. The dataset includes manual annotations of the left ventricular endocardium (${LV}_{endo}$), left ventricular epicardium (${LV}_{epi}$), and left atrium (LA) provided by cardiologists. Approximately half of these patients had a left ventricular ejection fraction less than 45$\%$, which is considered to indicate lesion risk. Among them, 19$\%$ of the images were rated as poor quality. While such images are typically excluded in conventional analyses, they were incorporated into the training and validation sets in this study to evaluate their influence on deep learning models. In addition, only end-diastolic (ED) and end-systolic (ES) frames with labeled information were used for training and validation.

\subsection{Results and discussion}
\subsubsection{EchoNet-Dynamic dataset}
This paper presents a systematic comparison of the performance of multiple segmentation models on the EchoNet-Dynamic dataset, and the results are presented in Table 1. The experimental findings indicate that traditional segmentation models, such as FCN \cite{25-long2015fully}, U-Net \cite{5-weng2021inet}, and PSP-Net \cite{26-zhao2017pyramid}, exhibit relatively consistent overall performance. However, their accuracy tends to be lower than that of the more sophisticated DeepLabv3+ architecture \cite{27-chen2018encoder}. This finding suggests that DeepLabv3+ is better at performing cardiac segmentation tasks, especially in terms of feature extraction and global context modeling. In addition, substantial advantages are observed for improved models, such as attention U-Net \cite{10-zhu2022attention}, U-Net++ \cite{28-zhou2018unet++}, and ResUnet \cite{29-diakogiannis2020resunet}, which outperform their predecessors, largely because of the incorporation of attention mechanisms and advanced network optimization techniques. In ES segmentation, these models display more vigorous boundary processing capabilities and are more precise in capturing complex edge information.

\begin{table}[htb]
    \caption{Comparison of experimental results on the EchoNet-Dynamic dataset.\label{tab:table1}}
    \centering
    \resizebox{0.5\textwidth}{!}{ 
    \begin{tabular}{p{2.0cm} p{1.2cm}<{\centering} p{1.2cm}<{\centering} p{1.2cm}<{\centering}}
        \toprule
        Method & Overall & ES & ED \\ 
        \midrule
        FCN & 91.89 & 90.36 & 92.86 \\
        U-Net & 91.94 & 90.39 & 92.91 \\
        DeepLabv3+ & 92.15 & 90.68 & 93.07 \\
        PSP-Net & 91.48 & 89.81 & 92.48 \\
        ResUnet++ & 92.21 & 91.18 & 92.85 \\
        ResUnet & 92.62 & 91.50 & 93.33 \\
        Attention U-Net & 92.44 & 91.31 & 93.15 \\
        U-Net++ & 92.76 & 91.42 & 93.60 \\
        EchoNet & 91.96 & 90.39 & 92.96 \\
        EchoGraphs & 92.10 & - & - \\
        VM-Unet & 92.84 & 91.60 & 93.61 \\
        LMaUNet & 92.89 & 91.59 & \textbf{93.70} \\
        Ours & \textbf{92.92} & \textbf{91.72} & 93.66 \\
        \bottomrule
    \end{tabular}
    }
    \renewcommand\arraystretch{1.5}
\end{table}

Furthermore, we compared models specifically designed for left ventricular segmentation, such as EchoNet \cite{24-ouyang2020video} and EchoGraphs \cite{31-thomas2022light}, and Mamba-based segmentation models, including VM-Unet \cite{19-ruan2024vm} and LMaUNet \cite{22-wang2024lkm}. The latest Mamba-based segmentation model showed significant potential and outperforms previous CNN-based and transformer-based models. Our model demonstrated substantial competitiveness among multiple evaluation metrics, particularly in terms of overall performance, where it achieved the highest accuracy of 92.92$\%$. For ES segmentation, our model achieved an 
accuracy
 of 91.72$\%$, surpassing that of other existing models. For ED segmentation, our model achieved an accuracy of 93.66$\%$, which is comparable to that of current state-of-the-art models. These results highlight that our method not only achieves exceptional performance in global cardiac contour extraction but also exhibits strong robustness and consistency across different cardiac phases, such as ES and ED. Overall, the experimental results indicate that the proposed model performs best in terms of overall segmentation accuracy and the detailed capture of key cardiac states. In addition, this model exhibits stable and efficient segmentation performance, even in the presence of complex changes in cardiac morphology.

\subsubsection{CAMUS dataset}
For the CAMUS dataset \cite{32-leclerc2019deep}, only the ES and ED frames containing annotation information were used for model training. Some data augmentation techniques commonly employed by mainstream models, such as rotation and denoising preprocessing operations, have been applied to mitigate the effects of common noise and deformation in ultrasound images. These techniques aim to increase the model's generalizability and robustness. Then, several state-of-the-art models were compared, and their performance was analyzed in multitarget segmentation tasks, specifically for the left ventricular endocardium, left ventricular epicardium, and left atrium.

First, a comparison was made with the GUDU model \cite{33-sfakianakis2023gudu}, which makes prediction via five trained U-Net architectures integrated with various data augmentation methods. While a direct comparison of ${LV}_{epi}$ segmentation is not feasible because of differences in the evaluation criteria, the proposed model outperforms GUDU in the segmentation tasks for ${LV}_{endo}$ and LA, where it achieves higher accuracy. Second, under the same data augmentation conditions and multitarget segmentation task, the proposed model was also compared with the following advanced models: ResUNet++ \cite{34-jha2019resunet++}, which is based on U-Net; DeepLabv3+ \cite{27-chen2018encoder}, which is based on a CNN; Attention-Unet \cite{10-zhu2022attention}, which combines attention and U-Net; SwinUNet \cite{35-cao2022swin}, which is based on a transformer; H2Former \cite{36-he2023h2former}, which integrates a CNN and transformer; and the latest Mamba-based LMaUNet \cite{22-wang2024lkm}.

\begin{table*}[htb]
    \caption{Comparison of experimental results on the CAMUS dataset.\label{tab:tabel2}}
    \centering
    \resizebox{0.92\textwidth}{!}{ 
    \begin{tabular}{p{2.0cm} p{1.5cm}<{\centering} p{1.5cm}<{\centering} p{1.5cm}<{\centering} p{1.5cm}<{\centering} p{1.5cm}<{\centering} p{1.5cm}<{\centering}}
        \toprule
        Method & ${LV}_{endo}$-ED & ${LV}_{endo}$-ES & ${LV}_{epi}$-ED & ${LV}_{epi}$-ES & LA-ED & LA-ES \\ 
        \midrule
        GUDU & 94.6 & 92.9 & - & - & 89.4 & 92.6 \\
        ResUNet++ & 93.10 & 91.31 & 83.23 & 84.36 & 88.20 & 90.90 \\
        DeepLabv3+ & 94.31 & 92.32 & 85.39 & 86.45 & \textbf{91.88} & \textbf{92.90} \\
        Attention-Unet & 92.77 & 89.56 & 82.86 & 82.84 & 88.22 & 89.93 \\
        SwinUNet & 92.43 & 89.54 & 81.89 & 82.10 & 85.96 & 89.43 \\
        H2Former & 94.15 & 92.44 & 85.72 & 86.55 & 89.88 & 92.07 \\
        LMaUNet & 94.47 & 92.34 & 86.44 & 86.69 & 89.82 & 91.58 \\
        Ours & \textbf{95.01} & \textbf{93.36} & \textbf{87.35} & \textbf{87.80} & 91.63 & 92.89 \\
        \bottomrule
    \end{tabular}
    }
    \renewcommand\arraystretch{1.5}
\end{table*}

The results indicate that SwinUNet demonstrates relatively poor performance, with the lowest segmentation accuracy for ${LV}_{endo}$, ${LV}_{epi}$, and LA. This may be attributed to the transformer's limitations in capturing details. Attention-Unet demonstrates improvements in segmentation accuracy across various targets, although the differences are not substantial. In contrast, models with CNN architectures, such as ResUNet++ and DeepLabv3+, benefit from the robust detail extraction capabilities of CNNs, which results in higher accuracy that aligns better with the high-precision requirements typical of medical image segmentation tasks. The H2Former model, which integrates a CNN and a transformer, outperforms ResUNet++ in all segmentation tasks and outperforms DeepLabv3+ for some targets. The LMaUNet model displays strong competitiveness in multitarget segmentation and confirms the effectiveness of the Mamba architecture in medical image segmentation.  

Ultimately, the proposed model achieves the best performance in the segmentation of ${LV}_{endo}$ and ${LV}_{epi}$ and ranks second only to DeepLabv3+ in LA segmentation. Notably, the model is outstanding in terms of ${LV}_{endo}$-ED, ${LV}_{endo}$-ES, ${LV}_{epi}$-ED, and ${LV}_{epi}$-ES segmentation, with the highest accuracies of 95.01, 93.36, 87.35, and 87.80, respectively. This finding highlights the superiority the proposed model in multitarget segmentation tasks. The segmentation accuracies for LA-ED and LA-ES are 91.63 and 92.89, respectively, second only to the accuracies of 91.88 and 92.90 achieved by DeepLabv3+. These results illustrate that the proposed model possesses strong robustness and generalization capabilities to address noise and deformation issues inherent in ultrasound images.

This paper presents a comparison of the segmentation results of various models on the CAMUS dataset, as illustrated in Figure 5. The first column displays the input image, the middle columns show the segmentation outputs from different models, and the last column presents the ground truth. To underscore the performance disparities among the models in intricate scenarios, a few challenging images were chosen for comparison. The segmentation results from the 2D apical two-chamber view are presented in the initial three rows. These images contain significant speckle noise around the left ventricle, which misleads certain models into incorrectly identifying it as the ventricle edge, leading to inaccurate segmentation. Many traditional models struggle to handle this noise effectively, failing to distinguish it from the actual ventricular boundary. The last three rows display the segmentation results from the 2D apical four-chamber view, which includes ED and ES frames. These images exhibit cases with severely blurred boundaries, where almost all competing models exhibit jitter and instability during segmentation. As a result, the boundary lines lack smoothness, and the segmentation accuracy is reduced. In contrast, the proposed method accurately delineates the true boundary of the ventricle even in such complex scenarios, and it displays superior robustness and stability. These visualization results highlight that the proposed method outperforms other methods in handling poor image quality, which provides more reliable segmentation outcomes.

\begin{figure*}[!htb]
\centering
\includegraphics[width=0.95\textwidth]{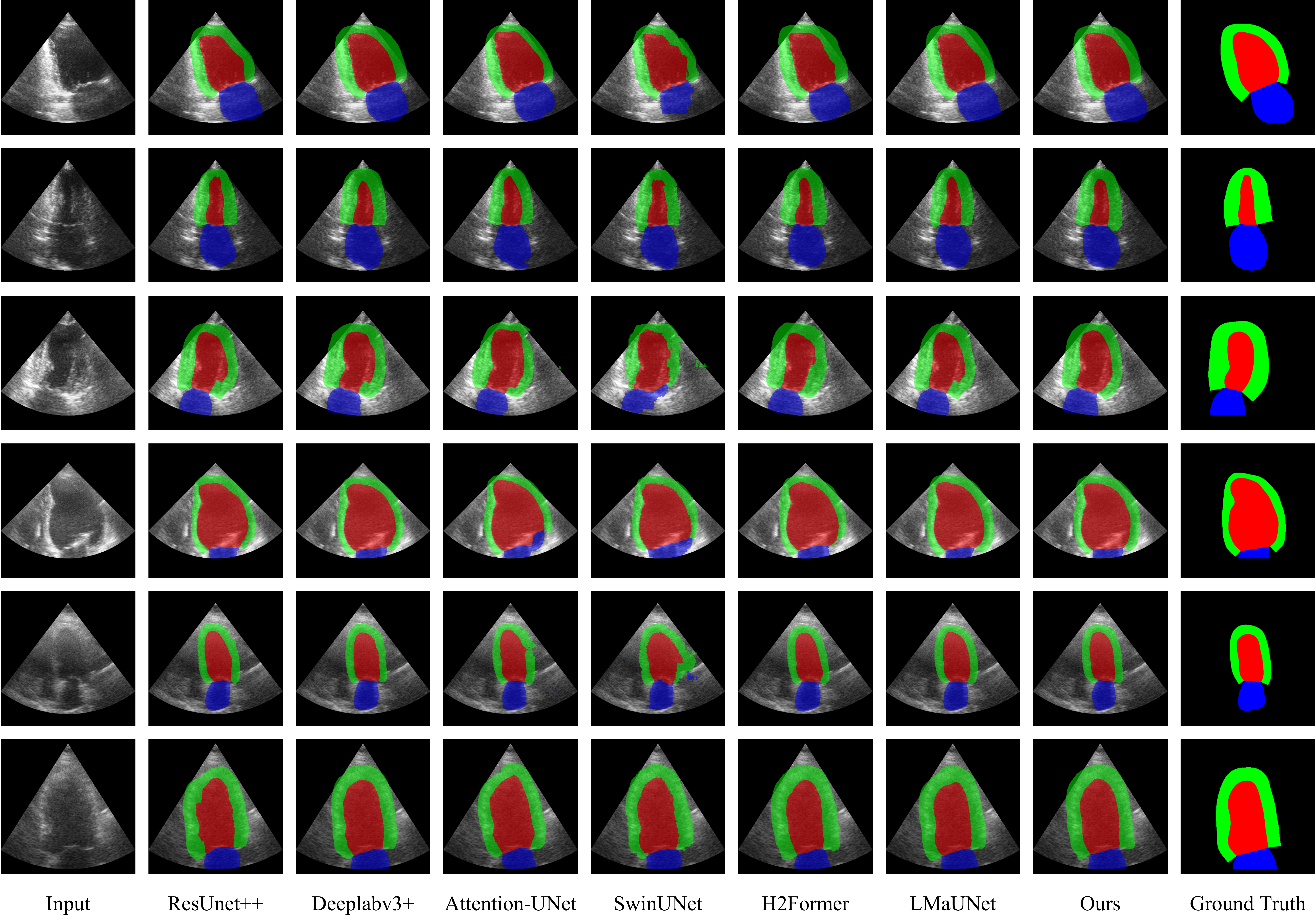}
\caption{Visual comparison of the segmentation results for the CAMUS dataset.}
\label{fig_5}
\end{figure*}

\subsubsection{Ablation experiment}
In the ablation experiment conducted in this work, the contribution of each model module to the overall performance of cardiac segmentation was systematically evaluated. The experiment was carried out on the EchoNet-Dynamic dataset, and the results are summarized in Table 3. When all the modules were integrated, the proposed model achieved the highest segmentation performance, with overall accuracies of 92.92$\%$, 91.72$\%$ at the ES stage, and 93.66$\%$ at the ED stage. These results demonstrate that the comprehensive integration of all the model modules considerably reinforces both the robustness and performance, which highlights exceptional feature extraction capabilities.

In contrast, the removal of certain modules negatively impacts the model's performance. For example, when the LMS block is removed and only auxiliary loss and multiscale feature fusion are employed, the overall accuracy decreases to 92.75$\%$, with performance values of 91.61$\%$ at the ES stage and 93.46$\%$ at the ED stage. When the LMS block is included but the auxiliary loss is omitted, the overall accuracy slightly decreases to 92.77$\%$, and the accuracy at the ED stage improves to 93.55$\%$. This finding highlights the positive influence of the LMS block on the processing of diastolic features. Further experimental analysis reveals that when the LMS block and auxiliary loss are used together, the overall accuracy increases to 92.80$\%$. The combination of these two modules improves the model's segmentation performance.


\begin{table}[htb]
    \caption{Ablation experiments on the EchoNet-dynamic dataset.\label{tab:tabel3}}
    \centering
    \renewcommand\arraystretch{1.5}  
    \resizebox{0.5\textwidth}{!}{
    \begin{tabular}{ccc>{\centering\arraybackslash}p{0.8cm}>{\centering\arraybackslash}p{0.8cm}>{\centering\arraybackslash}p{0.8cm}}
        \toprule
        LMS Block & Aux\_Loss & MSAA & overall & ES & ED \\ 
        \midrule
        \XSolidBrush & \checkmark & \checkmark & 92.75 & 91.61 & 93.46 \\
        \checkmark & \XSolidBrush & \checkmark & 92.77 & 91.52 & 93.55 \\
        \checkmark & \checkmark & \XSolidBrush & 92.80 & 91.62 & 93.53 \\
        \checkmark & \checkmark & \checkmark & \textbf{92.92} & \textbf{91.72} & \textbf{93.66} \\
        \bottomrule
    \end{tabular}}
\end{table}

Considering the results mentioned above, it is evident that the synergy between the modules, particularly when all modules are combined, optimizes the segmentation performance. The exclusion of specific modules directly reduces the model performance, indicating that every module plays an essential role in cardiac segmentation. This is especially apparent when processing images of different heart states, where the model displays notable robustness and comprehensiveness. This finding further emphasizes the importance of each component in achieving accurate and reliable segmentation outcomes.

\section{Conclusion and future work}
In this work, a U-shaped segmentation model based on multiscale vision Mamba feature fusion substantially improves the accuracy of echocardiographic segmentation tasks. The proposed model employs a residual block as an encoder and an LMS block as a decoder that effectively extracts the global and local features of the echocardiogram. The multiscale features and the global background were fused through the MSAA, while auxiliary losses were leveraged to increase the learning capacity of each layer's output. This approach facilitates the extraction of more representative features from complex medical images. Experiments conducted on the EchoNet-dynamic and CAMUS datasets demonstrate that the proposed model outperforms previous algorithms in segmentation accuracy for both single- and multisegmentation tasks in echocardiography.

In future studies, we intend to explore the advantages of three-dimensional (3D) reconstruction via echocardiography for assessing cardiac diseases. 3D reconstruction offers a more holistic perspective of cardiac structures, enabling improved visualization of anatomical relationships and functional dynamics. Our investigation focuses on algorithms for 3D reconstruction that efficiently leverage segmented data to generate detailed and precise heart models. This approach is expected to promote our comprehension of cardiac anatomy and improve diagnostic accuracy for patients with heart conditions. In addition, it will facilitate the formulation of diagnostic plans and further complete diagnostic capabilities.

\bibliographystyle{IEEEtran}
\bibliography{reference}

\begin{thebibliography}{10}
\providecommand{\url}[1]{#1}
\csname url@samestyle\endcsname
\providecommand{\newblock}{\relax}
\providecommand{\bibinfo}[2]{#2}
\providecommand{\BIBentrySTDinterwordspacing}{\spaceskip=0pt\relax}
\providecommand{\BIBentryALTinterwordstretchfactor}{4}
\providecommand{\BIBentryALTinterwordspacing}{\spaceskip=\fontdimen2\font plus
\BIBentryALTinterwordstretchfactor\fontdimen3\font minus \fontdimen4\font\relax}
\providecommand{\BIBforeignlanguage}[2]{{%
\expandafter\ifx\csname l@#1\endcsname\relax
\typeout{** WARNING: IEEEtran.bst: No hyphenation pattern has been}%
\typeout{** loaded for the language `#1'. Using the pattern for}%
\typeout{** the default language instead.}%
\else
\language=\csname l@#1\endcsname
\fi
#2}}
\providecommand{\BIBdecl}{\relax}
\BIBdecl

\bibitem{1-townsend2022epidemiology}
N.~Townsend, D.~Kazakiewicz, F.~Lucy~Wright, A.~Timmis, R.~Huculeci, A.~Torbica, C.~P. Gale, S.~Achenbach, F.~Weidinger, and P.~Vardas, ``Epidemiology of cardiovascular disease in europe,'' \emph{Nature Reviews Cardiology}, vol.~19, no.~2, pp. 133--143, 2022.

\bibitem{2-zhao2021epidemiological}
D.~Zhao, ``Epidemiological features of cardiovascular disease in asia,'' \emph{JACC: Asia}, vol.~1, no.~1, pp. 1--13, 2021.

\bibitem{3-halliday2021assessing}
B.~P. Halliday, R.~Senior, and D.~J. Pennell, ``Assessing left ventricular systolic function: from ejection fraction to strain analysis,'' \emph{European Heart Journal}, vol.~42, no.~7, pp. 789--797, 2021.

\bibitem{37-yildiz2020left}
M.~Yildiz, A.~A. Oktay, M.~H. Stewart, R.~V. Milani, H.~O. Ventura, and C.~J. Lavie, ``Left ventricular hypertrophy and hypertension,'' \emph{Progress in cardiovascular diseases}, vol.~63, no.~1, pp. 10--21, 2020.

\bibitem{40-nagueh2020left}
S.~F. Nagueh, ``Left ventricular diastolic function: understanding pathophysiology, diagnosis, and prognosis with echocardiography,'' \emph{JACC: Cardiovascular Imaging}, vol.~13, no. 1 Part 2, pp. 228--244, 2020.

\bibitem{38-pellikka2020guidelines}
P.~A. Pellikka, A.~Arruda-Olson, F.~A. Chaudhry, M.~H. Chen, J.~E. Marshall, T.~R. Porter, and S.~G. Sawada, ``Guidelines for performance, interpretation, and application of stress echocardiography in ischemic heart disease: from the american society of echocardiography,'' \emph{Journal of the American Society of Echocardiography}, vol.~33, no.~1, pp. 1--41, 2020.

\bibitem{39-pandian2023recommendations}
N.~G. Pandian, J.~K. Kim, J.~A. Arias-Godinez, G.~R. Marx, H.~I. Michelena, J.~C. Mohan, K.~O. Ogunyankin, R.~E. Ronderos, L.~E. Sade, A.~Sadeghpour \emph{et~al.}, ``Recommendations for the use of echocardiography in the evaluation of rheumatic heart disease: a report from the american society of echocardiography,'' \emph{Journal of the American Society of Echocardiography}, vol.~36, no.~1, pp. 3--28, 2023.

\bibitem{4-zamzmi2022real}
G.~Zamzmi, S.~Rajaraman, L.-Y. Hsu, V.~Sachdev, and S.~Antani, ``Real-time echocardiography image analysis and quantification of cardiac indices,'' \emph{Medical image analysis}, vol.~80, p. 102438, 2022.

\bibitem{5-weng2021inet}
W.~Weng and X.~Zhu, ``Inet: convolutional networks for biomedical image segmentation,'' \emph{Ieee Access}, vol.~9, pp. 16\,591--16\,603, 2021.

\bibitem{6-li2021survey}
Z.~Li, F.~Liu, W.~Yang, S.~Peng, and J.~Zhou, ``A survey of convolutional neural networks: analysis, applications, and prospects,'' \emph{IEEE transactions on neural networks and learning systems}, vol.~33, no.~12, pp. 6999--7019, 2021.

\bibitem{7-dosovitskiy2020vit}
A.~Dosovitskiy, L.~Beyer, A.~Kolesnikov, D.~Weissenborn, X.~Zhai, T.~Unterthiner, M.~Dehghani, M.~Minderer, G.~Heigold, S.~Gelly, J.~Uszkoreit, and N.~Houlsby, ``An image is worth 16x16 words: Transformers for image recognition at scale,'' \emph{ICLR}, 2021.

\bibitem{8-zhu2024vision}
L.~Zhu, B.~Liao, Q.~Zhang, X.~Wang, W.~Liu, and X.~Wang, ``Vision mamba: Efficient visual representation learning with bidirectional state space model,'' \emph{arXiv preprint arXiv:2401.09417}, 2024.

\bibitem{9-liu2021review}
X.~Liu, L.~Song, S.~Liu, and Y.~Zhang, ``A review of deep-learning-based medical image segmentation methods,'' \emph{Sustainability}, vol.~13, no.~3, p. 1224, 2021.

\bibitem{10-zhu2022attention}
Z.~Zhu, Y.~Yan, R.~Xu, Y.~Zi, and J.~Wang, ``Attention-unet: A deep learning approach for fast and accurate segmentation in medical imaging,'' \emph{Journal of Computer Science and Software Applications}, vol.~2, no.~4, pp. 24--31, 2022.

\bibitem{11-kushnure2021ms}
D.~T. Kushnure and S.~N. Talbar, ``Ms-unet: A multi-scale unet with feature recalibration approach for automatic liver and tumor segmentation in ct images,'' \emph{Computerized Medical Imaging and Graphics}, vol.~89, p. 101885, 2021.

\bibitem{12-zhang2020attention}
J.~Zhang, Z.~Jiang, J.~Dong, Y.~Hou, and B.~Liu, ``Attention gate resu-net for automatic mri brain tumor segmentation,'' \emph{IEEE Access}, vol.~8, pp. 58\,533--58\,545, 2020.

\bibitem{13-feng2020ssn}
R.~Feng, B.~Lei, W.~Wang, T.~Chen, J.~Chen, D.~Z. Chen, and J.~Wu, ``Ssn: A stair-shape network for real-time polyp segmentation in colonoscopy images,'' in \emph{2020 IEEE 17th International Symposium on Biomedical Imaging (ISBI)}.\hskip 1em plus 0.5em minus 0.4em\relax IEEE, 2020, pp. 225--229.

\bibitem{14-chen2024transunet}
J.~Chen, J.~Mei, X.~Li, Y.~Lu, Q.~Yu, Q.~Wei, X.~Luo, Y.~Xie, E.~Adeli, Y.~Wang \emph{et~al.}, ``Transunet: Rethinking the u-net architecture design for medical image segmentation through the lens of transformers,'' \emph{Medical Image Analysis}, p. 103280, 2024.

\bibitem{15-cao2022swin}
H.~Cao, Y.~Wang, J.~Chen, D.~Jiang, X.~Zhang, Q.~Tian, and M.~Wang, ``Swin-unet: Unet-like pure transformer for medical image segmentation,'' in \emph{European conference on computer vision}.\hskip 1em plus 0.5em minus 0.4em\relax Springer, 2022, pp. 205--218.

\bibitem{16-pham2024seunet}
T.-H. Pham, X.~Li, and K.-D. Nguyen, ``Seunet-trans: A simple yet effective unet-transformer model for medical image segmentation,'' \emph{IEEE Access}, 2024.

\bibitem{17-song2023tgdaunet}
P.~Song, J.~Li, H.~Fan, and L.~Fan, ``Tgdaunet: Transformer and gcnn based dual-branch attention unet for medical image segmentation,'' \emph{Computers in Biology and Medicine}, vol. 167, p. 107583, 2023.

\bibitem{18-bi2023bpat}
H.~Bi, C.~Cai, J.~Sun, Y.~Jiang, G.~Lu, H.~Shu, and X.~Ni, ``Bpat-unet: Boundary preserving assembled transformer unet for ultrasound thyroid nodule segmentation,'' \emph{Computer methods and programs in biomedicine}, vol. 238, p. 107614, 2023.

\bibitem{19-ruan2024vm}
J.~Ruan and S.~Xiang, ``Vm-unet: Vision mamba unet for medical image segmentation,'' \emph{arXiv preprint arXiv:2402.02491}, 2024.

\bibitem{20-liao2024lightm}
W.~Liao, Y.~Zhu, X.~Wang, C.~Pan, Y.~Wang, and L.~Ma, ``Lightm-unet: Mamba assists in lightweight unet for medical image segmentation,'' \emph{arXiv preprint arXiv:2403.05246}, 2024.

\bibitem{21-wang2024mamba}
Z.~Wang, J.-Q. Zheng, Y.~Zhang, G.~Cui, and L.~Li, ``Mamba-unet: Unet-like pure visual mamba for medical image segmentation,'' \emph{arXiv preprint arXiv:2402.05079}, 2024.

\bibitem{22-wang2024lkm}
J.~Wang, J.~Chen, D.~Chen, and J.~Wu, ``Lkm-unet: Large kernel vision mamba unet for medical image segmentation,'' in \emph{International Conference on Medical Image Computing and Computer-Assisted Intervention}.\hskip 1em plus 0.5em minus 0.4em\relax Springer, 2024, pp. 360--370.

\bibitem{23-wang2024weak}
Z.~Wang and C.~Ma, ``Weak-mamba-unet: Visual mamba makes cnn and vit work better for scribble-based medical image segmentation,'' \emph{arXiv preprint arXiv:2402.10887}, 2024.

\bibitem{30-ouyang2019echonet}
D.~Ouyang, B.~He, A.~Ghorbani, M.~P. Lungren, E.~A. Ashley, D.~H. Liang, and J.~Y. Zou, ``Echonet-dynamic: a large new cardiac motion video data resource for medical machine learning,'' in \emph{NeurIPS ML4H Workshop}, 2019, pp. 1--11.

\bibitem{32-leclerc2019deep}
S.~Leclerc, E.~Smistad, J.~Pedrosa, A.~{\O}stvik, F.~Cervenansky, F.~Espinosa, T.~Espeland, E.~A.~R. Berg, P.-M. Jodoin, T.~Grenier \emph{et~al.}, ``Deep learning for segmentation using an open large-scale dataset in 2d echocardiography,'' \emph{IEEE transactions on medical imaging}, vol.~38, no.~9, pp. 2198--2210, 2019.

\bibitem{25-long2015fully}
J.~Long, E.~Shelhamer, and T.~Darrell, ``Fully convolutional networks for semantic segmentation,'' in \emph{Proceedings of the IEEE conference on computer vision and pattern recognition}, 2015, pp. 3431--3440.

\bibitem{26-zhao2017pyramid}
H.~Zhao, J.~Shi, X.~Qi, X.~Wang, and J.~Jia, ``Pyramid scene parsing network,'' in \emph{Proceedings of the IEEE conference on computer vision and pattern recognition}, 2017, pp. 2881--2890.

\bibitem{27-chen2018encoder}
L.-C. Chen, Y.~Zhu, G.~Papandreou, F.~Schroff, and H.~Adam, ``Encoder-decoder with atrous separable convolution for semantic image segmentation,'' in \emph{Proceedings of the European conference on computer vision (ECCV)}, 2018, pp. 801--818.

\bibitem{28-zhou2018unet++}
Z.~Zhou, M.~M. Rahman~Siddiquee, N.~Tajbakhsh, and J.~Liang, ``Unet++: A nested u-net architecture for medical image segmentation,'' in \emph{Deep Learning in Medical Image Analysis and Multimodal Learning for Clinical Decision Support: 4th International Workshop, DLMIA 2018, and 8th International Workshop, ML-CDS 2018, Held in Conjunction with MICCAI 2018, Granada, Spain, September 20, 2018, Proceedings 4}.\hskip 1em plus 0.5em minus 0.4em\relax Springer, 2018, pp. 3--11.

\bibitem{29-diakogiannis2020resunet}
F.~I. Diakogiannis, F.~Waldner, P.~Caccetta, and C.~Wu, ``Resunet-a: A deep learning framework for semantic segmentation of remotely sensed data,'' \emph{ISPRS Journal of Photogrammetry and Remote Sensing}, vol. 162, pp. 94--114, 2020.

\bibitem{24-ouyang2020video}
D.~Ouyang, B.~He, A.~Ghorbani, N.~Yuan, J.~Ebinger, C.~P. Langlotz, P.~A. Heidenreich, R.~A. Harrington, D.~H. Liang, E.~A. Ashley \emph{et~al.}, ``Video-based ai for beat-to-beat assessment of cardiac function,'' \emph{Nature}, vol. 580, no. 7802, pp. 252--256, 2020.

\bibitem{31-thomas2022light}
S.~Thomas, A.~Gilbert, and G.~Ben-Yosef, ``Light-weight spatio-temporal graphs for segmentation and ejection fraction prediction in cardiac ultrasound,'' in \emph{International Conference on Medical Image Computing and Computer-Assisted Intervention}.\hskip 1em plus 0.5em minus 0.4em\relax Springer, 2022, pp. 380--390.

\bibitem{33-sfakianakis2023gudu}
C.~Sfakianakis, G.~Simantiris, and G.~Tziritas, ``Gudu: Geometrically-constrained ultrasound data augmentation in u-net for echocardiography semantic segmentation,'' \emph{Biomedical Signal Processing and Control}, vol.~82, p. 104557, 2023.

\bibitem{34-jha2019resunet++}
D.~Jha, P.~H. Smedsrud, M.~A. Riegler, D.~Johansen, T.~De~Lange, P.~Halvorsen, and H.~D. Johansen, ``Resunet++: An advanced architecture for medical image segmentation,'' in \emph{2019 IEEE international symposium on multimedia (ISM)}.\hskip 1em plus 0.5em minus 0.4em\relax IEEE, 2019, pp. 225--2255.

\bibitem{35-cao2022swin}
H.~Cao, Y.~Wang, J.~Chen, D.~Jiang, X.~Zhang, Q.~Tian, and M.~Wang, ``Swin-unet: Unet-like pure transformer for medical image segmentation,'' in \emph{European conference on computer vision}.\hskip 1em plus 0.5em minus 0.4em\relax Springer, 2022, pp. 205--218.

\bibitem{36-he2023h2former}
A.~He, K.~Wang, T.~Li, C.~Du, S.~Xia, and H.~Fu, ``H2former: An efficient hierarchical hybrid transformer for medical image segmentation,'' \emph{IEEE Transactions on Medical Imaging}, vol.~42, no.~9, pp. 2763--2775, 2023.

\end{thebibliography}

\end{document}